\documentstyle[twoside,fleqn,espcrc2]{article}

\newcommand{\AmS}{{\protect\the\textfont2
  A\kern-.1667em\lower.5ex\hbox{M}\kern-.125emS}}

\title{Finite Temperature QCD Interfaces Out of Equilibrium}

\author{Michael C. Ogilvie\address{Department of Physics, 
        Washington University, \\ 
        St. Louis, MO 63130, USA}%
        \thanks{This work was supported by the U.S. Department of Energy
		under Contract No. DE-FG02-91-ER40628.}
        }
       
\begin{document}

\begin{abstract}
The properties of interfaces in non-equilibrium situations
are studied by constructing a density matrix with a
space-dependent temperature.
The temperature gradient gives rise to new terms
in the equation for the order parameter.  
Surface terms
induced in effective actions by abrupt temperature changes
provide
a natural theoretical framework for understanding 
the occurence of both continuous and discontinuous behavior
in the order parameter.
Monte Carlo simulation of pure QCD shows both kinds of
interfacial behavior.
Perturbation theory predicts a universal
profile in the high temperature phase, which can be tested
by Monte Carlo simulation.
\end{abstract}

\maketitle

\section{INTRODUCTION}

	Lattice gauge theory is a well understood theoretical approach to the
equilibrium properties of finite temperature QCD, especially to bulk
thermodynamic properties. Experimentally, finite
temperature effects in QCD are of great interest for studies of
the early universe and heavy ion collisions, both non-equilibrium situations.
Much of the theoretical uncertainty is associated with the interface between
hadronic matter and quark-gluon matter. 

It is well established that computer simulations can give
information about equilibrium interfaces, including the
surface tension and the interfacial width. In principle,
information about transport coefficients can be obtained
from equilibrium correlation functions, although there
are technical difficulties.

An alternative approach is to examine the interfacial
behavior given basic information about the system at a given
time. For example, we might know that the average energy
density in a region of space has some given value, while
the rest of space has another. Given this information,
it is natural to contruct a density matrix which maximizes
the entropy subject to that constraint on the energy.

The local equilibrium formulation
of Zubarev \cite{Zub74}  constructs the time evolution of the density
matrix as successive changes in parameter fields like the local temperature
and the local velocity. The time evolution is understood as motion from one
local equibrium state to another.
In the case of a system at rest,
the density matrix has the form
\begin{equation}
\rho = \frac {1}{Z} exp ( - \int \beta (\vec{x} ) {\cal H} (\vec{x} )) 
\end{equation}
where H is the energy density operator. Such a temperature profile
can be imposed in a lattice simulation by noting that the inverse
temperature $\beta$ is given by $N_ta$, where $N_t$ is the temporal size
of the lattice in units of the lattice spacing a. The length of a
in physical units is specified by making the bare coupling
constants space-dependent. In principle, this can be done in QCD in
such a way that the spatial lattice spacing $a_s$ remains constant while
the temporal lattice spacing $a_t$ varies \cite{Kar82}; in practice this is
unnecessary in several cases of interest. In applications to cosmology, 
temperature differences may be small. This will certainly be the case
if the nucleation rate is large and
only small undercoolings are possible. In a first approximation to
the environment in a heavy ion collision, the temperature inside the region
of interest can be taken as roughly constant at a high temperature, and the
region outside as very cold. Monte Carlo studies \cite{mco92}
have shown
that  the Polyakov loop interfacial profile is not sensitive to the outer
region temperature, if it is sufficiently cold.

\section{BOUNDARY CONDITIONS AND SURFACE TERMS}

At the equilibrium point of a first-order phase
transtion, a stable interface between two phase can exist.
The interfacial profile for any order parameter is described by
a one-dimensional kink solution which interpolates between degenerate
minima of the effective potential.
Out of equilibrium, the minima are no longer degenerate.
In the traditional image of the ball rolling in an inverted potential,
the ball starts to roll down from one hilltop, and overshoots
the other. Thus, it is necessary to understand dissipative effects to
study the properties of a non-equilibrium interface.
As shown below, a temperature gradient is associated with dissipative
mechanisms that allow the existence of an interface in non-
equilibrium situations. An abrupt change in the temperature will
give rise to a localized dissipation at the interface. In this way,
surface terms are created in effective actions.

\subsection{Semiclassical Argument}

Consider a simple scalar field theory, and imagine that the
temperature T(z) is a given function, slowly varying over a mean
free path of all particles. Semiclassical arguments show that an
effective action for the one one-dimensional field theory is given
by
\begin{equation}
\frac{S_{eff}}{A} = \int dz \frac{1}{T} \left[ \frac{1}{2}
\left(\frac{d \phi}{dz}\right)^2 + V_{eff}(\phi,T)\right]
\end{equation}
where $\phi$ is the order parameter, A is the cross-sectional area
transverse to z and $V_{eff}$ is the finite temperature effective
potential. If $V_{eff}$ has two minima, the equation for the interface is
given by
\begin{equation}
- \frac{d^2 \phi}{d z^2} + ( \frac{1}{T} \frac{dT}{dz} )
\frac{d \phi}{dz} + \frac{\partial V_{eff}}{\partial \phi} = 0
\end{equation}
No stable interface can form at a constant temperature other than
the critical temperature, where the two minima are degenerate.
However, in the presence of a temperature gradient, this can occur,
due to the dissipative term in the effective action. There is a
kind of virial theorem, which relates the change in potential
energy density across the barrier to the dissipative term:
\begin{equation}
\Delta V =
\int dz ( \frac{1}{T} \frac{dT}{dz} )
\left[ \left( \frac{d \phi}{dz}\right)^2 -T \frac{\partial V_{eff}}{\partial T} \right]
\end{equation}
where $\Delta V_{eff}$ is the difference in the effective potential
between $z = + \infty$ and $z = - \infty$.
Dissipation takes place only in those regions where the temperature
is changing.

The semiclassical
analysis indicates that a localized temperature gradient will produce
a localized dissipation of energy.
It is easier, however, to analyize the effects of a sharp temperature
interface from a lattice point of view.
This corresponds precisely to the role of surface terms in
interfacial problems in condensed matter physics.

\subsection{Renormalization Group}

The renormalization group provide a simple explanation for the
occurence of surface terms \cite{Lip82a}. Suppose a sharp planar
interface at z=0 is introduced into some lattice model, with
couplings $J_1$ for $z<0$ and $J_2$ for $z>0$. Any real space renormalization
group transformation applied to this model will yield new bulk
couplings $J_1'$ and $J_2'$ away from the interface. At the interface, a
new set of couplings $J_s'$ will be generated naturally. In this
manner, even a microscopically sharp interface will naturally
generate surface terms in an effective action.

\subsection{Interfacial discontinuity}

At the critical temperature of a first order phase transition,
the order parameter changes continuously across an interface.
Out of equilibrium, an expanding bubble will evolve
toward a sharp temperature discontinuity \cite{Ign94}.
The order parameter may or may not exhibit
discontinuous behavior. 
It is more common in
condensed matter studies of interfacial phenomena to think of
bulk first order transitions as giving rise to
first-order interfacial transitions; the possibility
of second-order interfacial behavior
at a first-order bulk transition was first pointed out
in 1982 by Lipowsky \cite{Lip82b}. 

\begin {figure} [htb]
\bigskip
\bigskip
\bigskip
\bigskip
\bigskip
\bigskip
\bigskip
\bigskip
\bigskip
\bigskip
\bigskip
\bigskip
\bigskip
\bigskip
\bigskip
\includegraphics{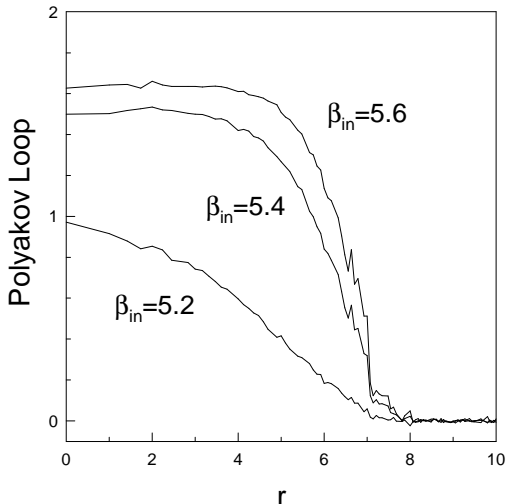}
\caption { Polyakov loop profile }
\label {fig: label means toungue? }
\end {figure}

\subsection{The interfacial profile in strong and weak coupling}

The strong coupling limit is a natural limit in which to study
the interfacial behavior of pure QCD, since the system
simplifies to a system of Polyakov loop spins.
Previous work using a mean field
theory applied to an SU(3) spin model derived from strong coupling
QCD shows that in a slab or semi-infinite geometry the surface
behavior is first order. This is in reasonable agreement with Monte
Carlo data \cite{mco93}.

Perturbative techniques can also be applied, using a variant of
the effective action developed for order-order interfaces 
in high-temperature QCD \cite{mco93,Bha91}.
This gives the interfacial profile for the Polyakov loop
on the high-temperature side of the deconfinement-confinement
interface.
The interfacial profile is a universal function of $gTz$.
However,
these methods are only useful in the high temperature regime, and
one cannot obtain a complete solution without an understanding
of the low temperature confining regime. 
Surface terms can be added to the effective action
to parametrize our ignorance; depending on the strength
of these terms, the profile may go continuously
to zero, or jump discontinuously to zero.

\section{MONTE CARLO RESULTS}

It is easy to simulate pure QCD at two different temperatures,
one above and one below the deconfinement transition.
The gas is at rest on both sides of the interface, and the
location of the wall is determined by temperature profile.
In the actual time evolution of the system, the interface
is moving, in such a way that total energy is conserved.

Figure 1 shows the Polyakov loop profiles 
for simulations of hot, deconfined spherical 
droplets of radius $R=7$ at $N_t=2$.
A spherical geometry was used, since it provides a detailed
interfacial profile.
The spherical surface introduces
corrections to the droplet free energy
which are proportional to the radius, but
this disadvantage is mild
compared to the rather coarse profile
provided by
a planar interface.
The coupling constant $\beta_{out}$ outside the bubble
was set at 1.0; no significant change in the profile
was seen at $\beta_{out}=3.0$, although boundary values changed slightly.
It was necessary to go to rather large values of
$\beta_{in}$, corresponding to high
temperatures, to obtain
an apparent sharp discontinuity in the Polyakov loop.
As shown in figure 2,
a mild discontinuity occurs when $\beta_{in}=5.6$ at $N_t=2$.
Given that $\beta_c =5.69$ at $N_t = 4$. this correspond to a
temperature of almost $ 2 T_c$. A fit to the continuum profile
\cite{mco93} for $\beta_{in}=5.6$
gives an excellent fit; however, the fitted value for gT is approximately
60\% higher than the perturbative value. 
A test of the perturbative
prediction for the profile at $N_t = 4$ or larger is indicated.

\end{document}